\documentclass[aps,pra,twocolumn,amsmath,amssymb,superscriptaddress]{revtex4}

\newcommand{\bra}[1]{\langle#1|}
\newcommand{\ket}[1]{|#1\rangle}

\usepackage{mathrsfs}
\usepackage[pdftex]{graphicx}

\begin{document}

\bibliographystyle{apsrev}

\title{Multi-walker discrete time quantum walks on arbitrary graphs, their properties, and their photonic implementation}

\author{Peter P. Rohde}
\email[]{dr.rohde@gmail.com}
\homepage{http://peterrohde.wordpress.com}
\affiliation{Max-Planck Institute for the Science of Light, Erlangen, Germany}
\affiliation{Centre for Quantum Computer Technology, University of Queensland, Australia}

\author{Andreas Schreiber}
\affiliation{Max-Planck Institute for the Science of Light, Erlangen, Germany}

\author{Martin {\v S}tefa{\v n}{\' a}k} \author{Igor Jex}
\affiliation{Department of Physics, Faculty of Nuclear Sciences and Physical Engineering, Czech Technical University in Prague, Czech Republic}

\author{Christine Silberhorn}
\affiliation{Max-Planck Institute for the Science of Light, Erlangen, Germany}
\affiliation{University of Paderborn, Paderborn, Germany}


\begin{abstract}
Quantum walks have emerged as an interesting alternative to the usual circuit model for quantum computing. While still universal for quantum computing, the quantum walk model has very different physical requirements, which lends itself more naturally to some physical implementations, such as linear optics. Numerous authors have considered walks with one or two walkers, on one dimensional graphs, and several experimental demonstrations have been performed. In this paper we discuss generalizing the model of discrete time quantum walks to the case of an arbitrary number of walkers acting on arbitrary graph structures. We present a formalism which allows for analysis of such situations, and several example scenarios for how our techniques can be applied. We consider the most important features of quantum walks -- measurement, distinguishability, characterization, and the distinction between classical and quantum interference. We also discuss the potential for physical implementation in the context of linear optics, which is of relevance to present day experiments.
\end{abstract}

\date{\today}

\frenchspacing

\maketitle


\section{Introduction}

Quantum walks \cite{bib:ADZ} offer an alternate approach to implementing quantum algorithms \cite{bib:AAKV} compared to the typical circuit \cite{bib:NielsenChuang00} or cluster state \cite{bib:Raussendorf01,bib:Raussendorf03} models for quantum algorithms. Indeed, quantum walks have been successfully applied to database search \cite{bib:SKW} and graph isomorphism \cite{bib:DW} problems. More recently, the universality of quantum walks has been shown in the case of continuous \cite{bib:Childs09} and discrete \cite{bib:Lovett10} systems. A review of both continuous and discrete walks can be found in Ref. \cite{bib:Kempe08}.

For more than a decade the quantum walk was a very fruitful yet theoretical concept. However, experiments with single optically trapped atoms \cite{bib:KFChSAMW} have successfuly demonstrated that quantum walks can be realized in the laboratory. A quantum walk on trapped ions \cite{bib:SMSGEHS} has been realized. Various successful realizations of a quantum walk based on quantum optics implementation followed.

Indeed, photonic  discrete time quantum walks were experimentally demonstrated by Schreiber et al. \cite{bib:Schreiber10} using weak coherent light and by Broome et. al \cite{bib:Broome10} using single photons produced via parametric down-conversion (PDC). The first experiments investigating the impact of correlations in quantum walks with two walkers were done in the continuous time model: Bromberg et. al \cite{bib:Bromberg09} showed correlations by classical intensity correlation measurements and Peruzzo et. al \cite{bib:Peruzzo10} used coincidence measurements of two photons of a PDC source, both with integrated wave guide devices.

Most of the studies to date have focused on walks with a single quantum particle. Nevertheless, some works have considered the effect of multiple walkers on a line. Indeed, quantum walks involving more particles unlock additional possibilities which are unavailable in classical random walks, as the quantum particles can be entangled or they can be indistinguishable fermions or bosons. In Ref. \cite{bib:VenegasAndraca09} the situation with two perfectly correlated walkers is considered, while in Refs. \cite{bib:Omar06,bib:Pathak07} the two walker situation is considered with separable and initially entangled walkers. The meeting problem in this model has been analyzed in Ref. \cite{bib:Stefanak06}. Additionally, Ref. \cite{bib:Goyal10} considers many-body entanglement using one dimensional walks.

In this paper we present the following: (1) a framework for multi-walker walks on arbitrary graph structures; (2) the difference between distinguishable and indistinguishable walkers, and the issue of walker correlations; (3) the characterization of quantum walks; (4) the effects of measurement on multi-walker systems in the context of both distinguishable and indistinguishable walkers; and, (5) we turn our attention to the example of a two photon linear walk with both distinguishable and indistinguishable walkers, which is of relevance to present day experiments. Our work builds on previous work on quantum walks by introducing techniques for modeling multiple walkers and arbitrary graph structures. Until now there has only been very limited consideration of multi-walker systems. We do this by introducing a \emph{walker operator} technique for describing such systems. In the context of the multiple walker scenario we discuss the main important issues surrounding measurement and characterization, and present a detailed discussion of the special case of photonic implementation.

The archetypal discrete time quantum walk consists of two building blocks -- a \emph{coin} operator and a \emph{step} operator. This terminology arises historically in accordance to models of discrete classical random walks. For consistency we kept the formalism, although the two operators can be merged into a single \emph{evolution operator}. We define a basis of states of the form $\ket{x,c}$, where $x$ denotes a \emph{position} state of a \emph{walker}, and $c$ denotes a value of the \emph{coin}. The position state could literally refer to the physical position of a particle, as in the case of most photonic implementations, or it could refer more abstractly to the occupation of a level in some higher dimensional degree of freedom. The coin is essentially an ancillary parameter which is used by the step operator to decide how to propagate the walker. The simplest example of a quantum walk is the one dimensional case with a single walker, where the positions $x$ represent positions along a one-dimensional lattice graph. A common choice of coin and step operators are,
\begin{eqnarray} \label{eq:linear_graph_one_walker}
H\ket{x,\pm 1} &=& (\ket{x,-1} \pm \ket{x,1})/\sqrt{2} \nonumber \\
S\ket{x,c} &=& \ket{x+c,c}
\end{eqnarray}
where $c \in \{-1,1\}$ and $x\in \{\mathbb{Z}\}$. $H$ is the Hadamard coin, and $S$ is the nearest neighbor step operation. Thus, the coin operator affects only the coin state and leaves the position state unchanged, whereas the step operator determines the evolution of the position state, as a function of the value of the coin, leaving the coin state unchanged. This particular choice of coin and step operators can be optically implemented using 50/50 beamsplitters. After $n$ rounds of the quantum walk, the evolution is of the form $(SH)^n$. At each step the reach of the position state expands outwards across the line. Additionally, because the position states propagate in both directions, interference between the position states takes place. This is the defining feature that differentiates the quantum walk from a classical random walk \footnote{The analogous classical random walk is that of a ball propagating down a triangular (Galton) pin-board, where, at each pin, the ball propagates either to the left or to the right with 50\% probability. As the balls do not undergo interference, the propagation results in classical Bernoulli statistics at the outputs.}.

In the above description of a quantum walk we have made a specific choice of coin and step operators. Of course, we have great flexibility in choosing these operators. In fact, the algorithm implemented by a quantum walk is characterized entirely by this choice. For example, we could use a biased coin, which can optically be implemented using a beamsplitter with reflectivity $\eta^2 \neq 1/2$, or a step operator with, for example, periodic boundary conditions, \mbox{$S\ket{x,c} = \ket{(x+c)\,\mathrm{mod}\, N,c}$}, where $N$ is the number of distinct position states. Not only can we choose arbitrary operators, but, in general, the choice can vary at each step of the walk. In the following sections we will describe techniques for analyzing multi-walker walks on arbitrary graphs, and some of their properties.


\section{Multi-walker quantum walks on arbitrary graphs} \label{sec:multi_walker}

We now turn our attention to generalizing the quantum walk to the case of multiple walkers and arbitrary graph structures. This is motivated by the increasing complexity of the system as the number of walkers increases. Additionally, algorithms are emerging relying on multiple walkers \cite{bib:Gamble10} and higher dimensional graph structures \cite{bib:Mackay02}. First let us develop a notation for representing multi-walker states. We begin by introducing the \emph{walker operator}, $w(x,c)^\dag$, which creates a walker at position $x$ with coin value $c$. We also define the \emph{empty walker state}, $\ket{0}$, which is the state of the system without any walkers present. In an optical context the walker operator corresponds directly to the photon creation operator, and the empty walker state is the vacuum state. The walker operator has the following effect,
\begin{equation}
w(x,c)^\dag\ket{0} = \ket{x,c}
\end{equation}
and all walker operators commute. An arbitrary initial state of a quantum walk system with multiple walkers can be represented as combinations of different walker operators.

Next we introduce the \emph{walker graph}, which captures the available position states, and the relationship between them. In general these graphs will be bidirectional, with weighted edges. Some examples are illustrated in Figs. \ref{fig:graph_linear} and \ref{fig:graph_lattice}. We define unitary operators $A^{(x)}$, which act on the \emph{neighborhood} of $x$. The neighborhood of position state $i$ is denoted $n_i$ and consists of $i$ and its immediate neighbors with outgoing directed edges.

Now we define the coin and step operators as follows,
\begin{eqnarray} \label{eq:general_coin_step_pd}
&C:& \, w(x,c)^\dag \mapsto \sum_{j\in n_x} A_{cj}^{(x)} w(x,j)^\dag \nonumber \\
&S:& \, w(x,j)^\dag \mapsto w(j,x)^\dag
\end{eqnarray}
where $A$ is unitary for all $x$. Note that the definition of the step operator in this case is not quite analogous to the simple example presented in Eq. \ref{eq:linear_graph_one_walker}. There the step operator updates the position while leaving the coin state unchanged, whereas in the above formalism the step operator is a permutation operator that changes both the coin \emph{and} the position. This slight change in the formalism is necessary to ensure unitarity of the step operator. For example, if in analogy to Eq. \ref{eq:linear_graph_one_walker} our step operator implemented the transformation $w(x,j)^\dag \to w(j,j)^\dag$, the operation would not be reversible.

The interpretation of the coin state varies at different points in the evolution of the walker. Prior to the coin operator, the coin value tells us the \emph{source} of the walker (i.e. its previous position state), whereas after the coin operator it tells us the \emph{destination} of the walker (i.e. its next position state). This is consistent with the definition of the linear walk in Eq. \ref{eq:linear_graph_one_walker}. Thus the coin operator puts the walker into a superposition of destination states within the neighborhood, and the step operator updates the walker to the new destination while replacing the source state with the previous position state. In general, time-dependent coins are allowed, and the evolution of the system will be of the form $\prod_i S C_i$, where $C_i$ is the coin operator at time step $i$.

This formalism maps a position to a destination, with memory of the previous state. In general, arbitrary graphs will have varying degree, i.e. different neighborhoods contain different numbers of vertices. Thus in general we cannot use a single coin to describe the dynamics of the entire graph. Instead, unique $A^{(x)}$ operators must be defined for every neighborhood. This differs from our initial example of a walk on a line, where the degree is constant and a single coin can characterize the entire system.

Importantly, in the above formalism we have implicitly assumed that the coin is independent of the number of walkers. For example, in the optical context one might imagine that the coin is based on a cross-Kerr effect, in which case the coin is dependent on the number of walkers. We implicitly rule out these kinds of effects such that, for example, in the optical case (to be discussed in detail in Sec. \ref{sec:optics}), propagation is defined entirely by linear elements -- beamsplitters and phase shifters.

Using this formalism the walk is characterized completely by (1) the initial state of the system, and (2) the unitary coin matrices of the graph. Additionally, the evolution of a particular walker is characterized entirely by its neighborhood in the graph. The evolution of the system, and the algorithm implemented by it, are specified solely by these parameters.


\subsection{Examples of graphs} \label{sec:examples_graphs}


\subsubsection{One-dimensional lattice graph}

\begin{figure}[!htb]
\includegraphics[width=0.47\textwidth]{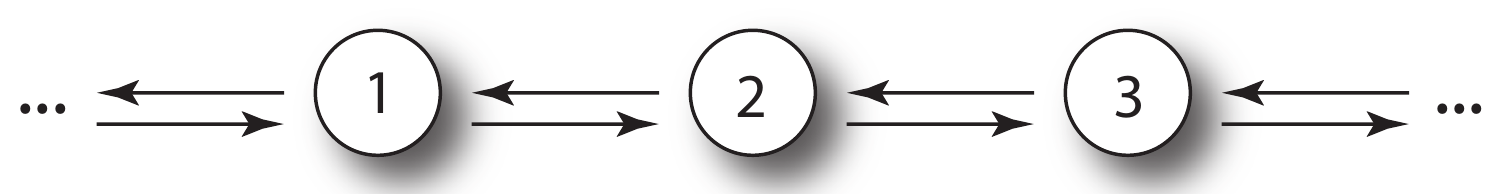}
\caption{Position graph for a one dimensional linear walk.} \label{fig:graph_linear}
\end{figure}

As a first simple example we will step backwards to the simple example of walkers on an infinite one-dimensional lattice graph. This graph is already well understood, so we will simply demonstrate in this section that the general formalism we have presented reduces to the dynamics presented earlier in Eq. \ref{eq:linear_graph_one_walker}. However our formalism accommodates for arbitrary number of walkers rather than just one. The position graph for the linear walk is shown in Fig. \ref{fig:graph_linear}.

To model this graph in our general formalism, we only have to chose appropriate operators $A^{(x)}$. For this graph we choose the matrix elements of the operators as follows,
\begin{eqnarray}
A^{(x)}_{n_x(1),n_x(1)} &=& 1/\sqrt{2} \,\,\,\, \forall \,\, x \nonumber \\
A^{(x)}_{n_x(1),n_x(2)} &=& 1/\sqrt{2} \,\,\,\, \forall \,\, x  \nonumber \\
A^{(x)}_{n_x(2),n_x(1)} &=& 1/\sqrt{2} \,\,\,\, \forall \,\, x  \nonumber \\
A^{(x)}_{n_x(2),n_x(2)} &=& -1/\sqrt{2} \,\,\,\, \forall \,\, x
\end{eqnarray}
where $n_x(i)$ is the $i$th neighbor of $x$ -- in this case there are only two. Thus the operator $A^{(x)}$ is the Hadamard operator, acting on the subspace spanned by $n_x$, in analogy to the linear system presented earlier. Note that we have made the coin operator the same for all values of $x$, which is legitimate when the line is infinitely long (i.e. without boundary conditions).


\subsubsection{Two dimensional lattice graph}

\begin{figure}[!htb]
\includegraphics[width=0.38\textwidth]{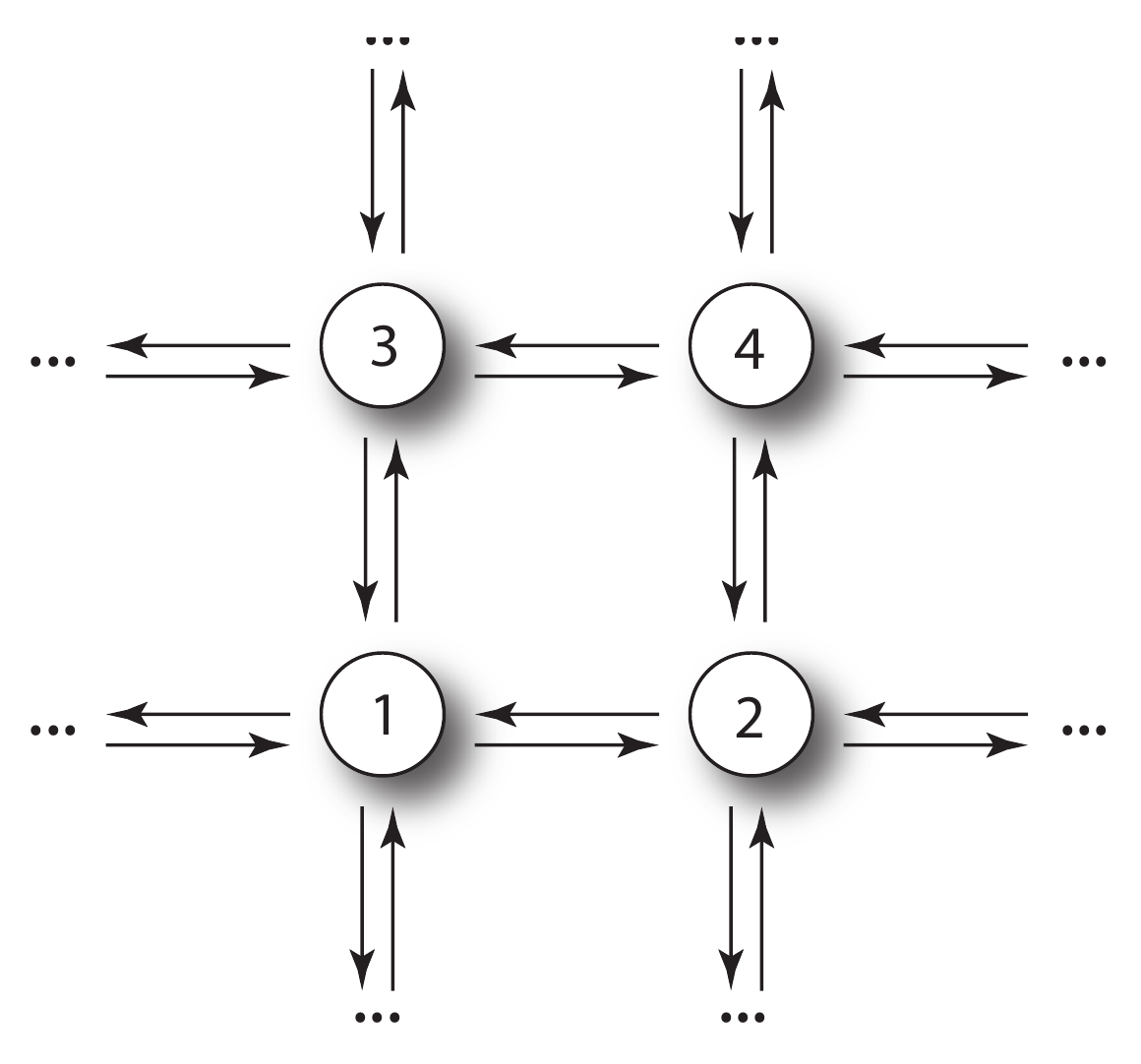}
\caption{Position graph for a walk on a two dimensional lattice.} \label{fig:graph_lattice}
\end{figure}

Next we turn our attention to a simple two dimensional example -- a two dimensional regular lattice. The position graph is shown in Fig. \ref{fig:graph_lattice}. In this example we will let the operators $A^{(x)}$ be the same for all values of $x$. Again, in general this would not apply with boundary conditions, but on an infinite lattice this is a valid assumption.

An example of a coin on this graph is the four level Grover coin, given by
\begin{equation} \label{eq:grover_coin}
G_4 = \frac{1}{2} \left( \begin{array}{cccc}
-1 & 1 & 1 & 1 \\
1 & -1 & 1 & 1 \\
1 & 1 & -1 & 1 \\
1 & 1 & 1 & -1 \\
\end{array} \right)
\end{equation}
which can be generalized to higher dimensions. This coin is essentially a generalization of the Hadamard coin to a four level system \footnote{Other generalizations of the Hadamard coin to four levels are possible, e.g. the four level discrete fourier transform (DFT) coin.}, and for a given position, defines the mapping between the old coin value and the new coin value.


\subsection{Distinguishable versus indistinguishable walkers}

In the above discussion we have implicitly assumed that walkers with the same position and coin values are indistinguishable, i.e. if we have two walkers in a given position and coin state, measurement cannot determine which walker is which. Should we want to model distinguishable walkers we can introduce a new parameter which identifies a given walker from the others. In this case we redefine our walker operator as
\begin{equation}
w(x,c,n)^\dag
\end{equation}
where $n$ is a new parameter which identifies which walker the operator corresponds to. In an optical context this could, for example, identify in which spatio-temporal mode or polarization the photon is in. We now have the property that $\bra{0}w(x,c,n)w(x,c,m)^\dag\ket{0}=\delta_{m,n}$. In the following sections we will primarily consider the situation of indistinguishable walkers, since this is the interesting case where the walkers interfere with one another. However, in Sec. \ref{sec:int_indep} we will consider the two cases in the optical context and examine the difference in the interference patterns.


\subsection{Characterizing quantum walks} \label{sec:char_walks}

A completed quantum walk, up to phases, can be fully characterized by a joint probability distribution (JPD) \cite{bib:Omar06,bib:Pathak07}. The JPD has dimension $n$, where $n$ is the number of walkers, each degree of freedom represents the position of a given walker, and each element in the JPD is the joint (i.e. coincidence) probability of measuring a walker at each of the corresponding positions. We define \mbox{$P_{i_1,i_2,\dots,i_n}$} to be the probability that a walker is found at each of the positions $i_1,i_2,\dots,i_n$ (an $n$-fold coincidence probability). Optically this would simply be implemented by placing photo-detectors at each position and measuring joint detection events.

The number of terms in the JPD is $O(N^n)$, where $N$ is the number of distinct position states. This implies that fully characterizing the statistics of a multi-walker quantum walk is infeasible with large numbers of walkers, necessitating the development of algorithm specific metrics for characterizing the system which do not require exponentially large numbers of measurements. This scaling raises the interesting side-thought as to how the computational power of a quantum walk scales with the number of walkers -- with a single walker the number of parameters characterizing the system is $O(N)$, whereas for $n>1$ the number of parameters grows exponentially with $n$. The JPD can be used to derive any output statistics of interest.

As an example, in the \emph{meeting problem} we are interested in the probability that all walkers are measured at the same position, which is given by $\mathrm{tr}(P)$. The meeting problem is one of the standard parameters discussed in connection with classical walks involving multiple particles \cite{bib:Hughes95}. The meeting probability quantifies the possibility the two walkers meet. In the quantum domain the meeting probability quantifies the analogous property but takes also into account additional quantum properties of the walkers. Hence it is natural to expect quite a different behavior of the meeting probability when compared to the classical case. The effects of walk topology and indistinguishability of the walker can be nicely represented. On the short-time scale, quantum features result in strong increase of the meeting probability for indistinguishable bosons, an effect familiar to Hong-Ou-Mandel interference. Moreover, the differences in the meeting probability can be found also in the asymptotic regime. However, the quantum effects on the long-time scale are much more subtle, as discussed in detail in Ref. \cite{bib:Stefanak06}.


\subsection{Measurement} \label{sec:measurement}

We now investigate the effects of measurement on a multi-walker system. We will focus on the case of two walkers, although this could be easily generalized to arbitrary numbers of walkers. Consider a two-walker system on an arbitrary graph with the first walker initialized into position $x$ and the second into position $y$. Then following $n$ applications of coin and step operators, $U=(SC)^n$, we obtain
\begin{equation}
w(x)^\dag w(y)^\dag \mapsto \Big(\sum_i U_{x,i} w(i)^\dag\Big)\Big(\sum_j U_{y,j} w(j)^\dag\Big)
\end{equation}
where we have ignored coin values for simplicity. Note that although this state is written in a factorizeable form it is nonetheless entangled, since it cannot be rewritten in a form which is a product of the walkers at different positions. Next, suppose we perform a measurement and detect a single walker at position $m$. Then the state will be projected into,
\begin{equation}
w(m)^\dag \sum_{i\neq m} (U_{x,m}U_{y,i} + U_{y,m}U_{x,i})w(i)^\dag
\end{equation}
Thus, upon measurement the factorizeable nature of the state is lost, and our resulting state contains cross-combinations from the different terms in the initially factorizeable state. Therefore the behavior of a multi-walker system cannot, in general, be characterized in terms of single walker systems post-measurement.

In an optical context this corresponds to preparing photons at positions $x$ and $y$, allowing them to propagate through a linear optics network (beamsplitters and phase-shifters) characterized by the unitary matrix $U$, and then measuring a single photon at position $m$.

We could also characterize the system in terms of the coincidence probability, which, for two walkers, is given by \cite{bib:Bromberg09}
\begin{equation}
P_{m,n} = |U_{x,m}U_{y,n} + U_{x,n}U_{y,m}|^2 \,\,\, \forall \,\,\, m\neq n
\end{equation}
Again, in an optical context this equates to measuring a photon at both positions $m$ and $n$ -- a standard coincidence measurement.

The above discussion applies for indistinguishable walkers. In the case of distinguishable walkers the state, upon measurement, is projected onto
\begin{equation}
w(m,1)^\dag U_{x,m} \Big(\sum_j U_{y,j} w(j,2)^\dag\Big)
\end{equation}
where $1$ and $2$ denote the classically distinguishable walkers and we have measured walker $1$. Similarly, the coincidence probability for distinguishable walkers is given by
\begin{equation} \label{eq:dist_coinc}
P_{m,n} = |U_{x,m} U_{y,n}|^2
\end{equation}
where we have measured the first walker at position $m$ and the second at $n$. This expression contains no interference terms, as expected. It is evident from Eq. \ref{eq:dist_coinc} that for distinguishable walkers the coincidence probability has an uncorrelated form, i.e. it can be represented as the product of two independent functions. Hence the coincidence statistics for distinguishable walkers can be characterized in terms of single walker walks. This will be discussed further in the optical context in Sec. \ref{sec:int_indep}.

An interesting observation is that when both walkers are initialized into the same position state, $x=y$, the coincidence probability $P_{m,n}$ is of the same form between the distinguishable and indistinguishable scenarios \footnote{The probability $P_{m,n}$ is the same for the distinguishable and indistinguishable cases, up to a constant factor of $4$ owing to the differing normalization conditions.}. This implies that we cannot expect a quantum speedup over the single walker case by preparing multiple walkers into the same initial state.

\section{Application to quantum optics: Two photon walk on a line} \label{sec:optics}


\subsection{The two photon beamsplitter coin}

We now turn our attention to the illustrative example of the two photon, one dimensional quantum walk. Suppose we have a system containing exactly two photons (walkers), and the propagation of the quantum walk is performed by beamsplitters. Then each input to a beamsplitter will have 0, 1 or 2 photons. We can define the basis states for the input and output of the beamsplitter as
$\ket{00}$, $\ket{01}$, $\ket{10}$, $\ket{11}$, $\ket{20}$ and $\ket{02}$, where $\ket{mn}$ is the two mode state with $m$ photons in the first mode and $n$ photons in the second. For now we assume the incident photons are indistinguishable. Acting on this basis, the beamsplitter transformation can be represented in matrix form,
\begin{equation} \label{eq:BS_transformation}
B_2 = \left( \begin{array}{c|cc|ccc}
1 & 0 & 0 & 0 & 0 & 0 \\
\hline
0 & -\frac{1}{\sqrt{2}} & \frac{1}{\sqrt{2}} & 0 & 0 & 0 \\
0 & \frac{1}{\sqrt{2}} & \frac{1}{\sqrt{2}} & 0 & 0 & 0 \\
\hline
0 & 0 & 0 & 0 & \frac{1}{\sqrt{2}} & -\frac{1}{\sqrt{2}} \\
0 & 0 & 0 & \frac{1}{\sqrt{2}} & \frac{1}{2} & \frac{1}{2} \\
0 & 0 & 0 & -\frac{1}{\sqrt{2}} & \frac{1}{2} & \frac{1}{2} \\
\end{array} \right)
\end{equation}
This follows directly from the beamsplitter equations of motion, $a^\dag\to (a^\dag+b^\dag)/\sqrt{2}$ and $b^\dag\to (a^\dag-b^\dag)/\sqrt{2}$ (using the asymmetric beamsplitter convention), where $a$ and $b$ denote the two input/output modes. The matrix is block diagonal, $B_2 = H_0 \oplus H_1 \oplus H_2$, with each of the blocks representing the different photon number sub-spaces: the zero photon block implements the trivial identity operation; the single photon block implements the usual Hadamard coin; and, the two photon block implements an unbalanced generalization of the Hadamard coin, with its top-left element zero, corresponding to Hong-Ou-Mandel (HOM) \cite{bib:HOM87} interference between the incident photons. In the single walker case, the beamsplitter coin will consist of just the first two blocks. Importantly, the $H_0$ and $H_1$ blocks  can be simulated using coherent light, whereas the $H_2$ block cannot be, as it contains HOM interference terms. This differentiates the single and two walker scenarios.

\begin{figure}[!htb]
\includegraphics[width=0.3\textwidth]{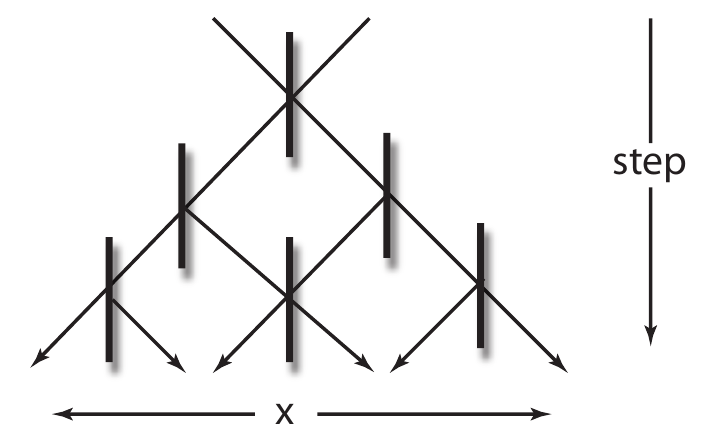}
\caption{A simple optical 3 level quantum walk on a line.} \label{fig:pyramid}
\end{figure}

Experimentally, the beamsplitter simultaneously implements the coin and step operators. That is, it simultaneously flips a quantum mechanical coin and subsequently propagates the physical state spatially. A simple optical quantum walk using beamsplitter coins is illustrated in Fig. \ref{fig:pyramid}. This configuration is ideally suited to implementation using integrated waveguides, as has been demonstrated in Ref. \cite{bib:Peruzzo10}.

\subsection{Photon statistics for the two photon quantum walk}

\begin{figure*}[!htb]
\includegraphics[width=0.8\textwidth]{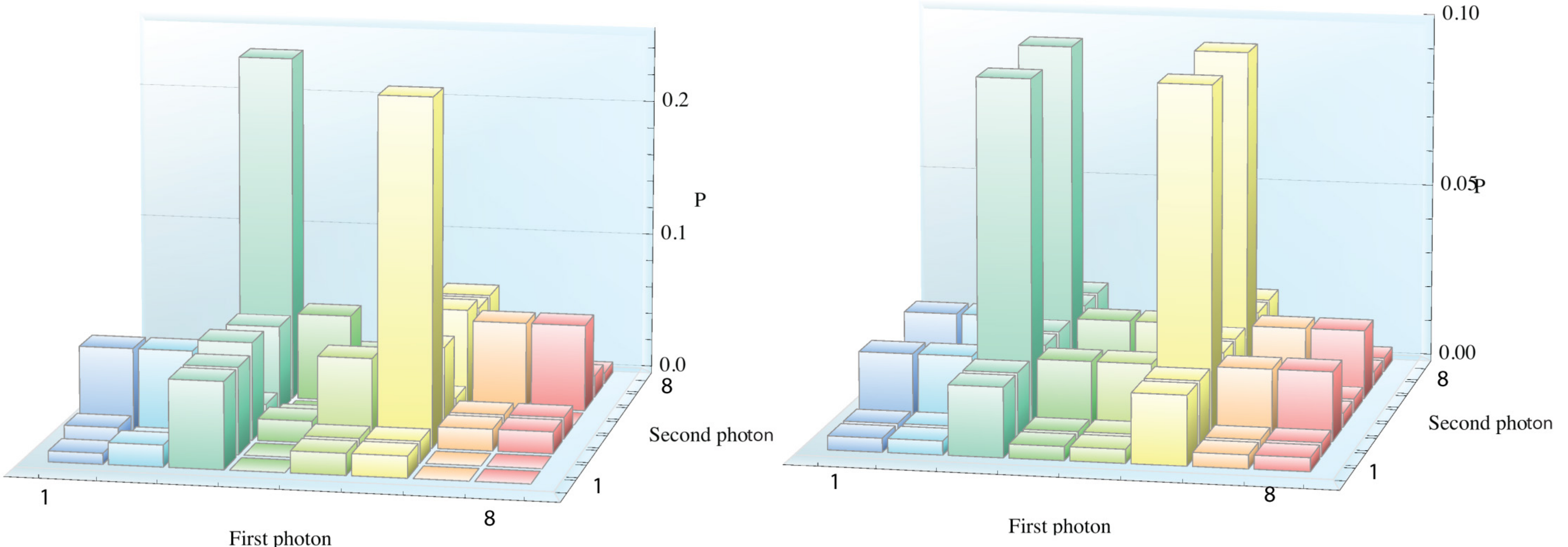}
\caption{(Color online). Two photon quantum walk with 4 levels. (left) Joint probability distribution with two indistinguishable walkers. (right) Joint probability distribution with two distinguishable walkers. Note that in the case of distinguishable walkers the coincidence matrix is uncorrelated and can be obtained by multiplying the single photon statistics along one axis by the single photon statistics along the other axis. In the case of indistinguishable walkers the matrix is correlated, which is characteristic of a quantum behavior.} \label{fig:theory_plots_4_levels}
\end{figure*}

\begin{figure*}[!htb]
\includegraphics[width=\textwidth]{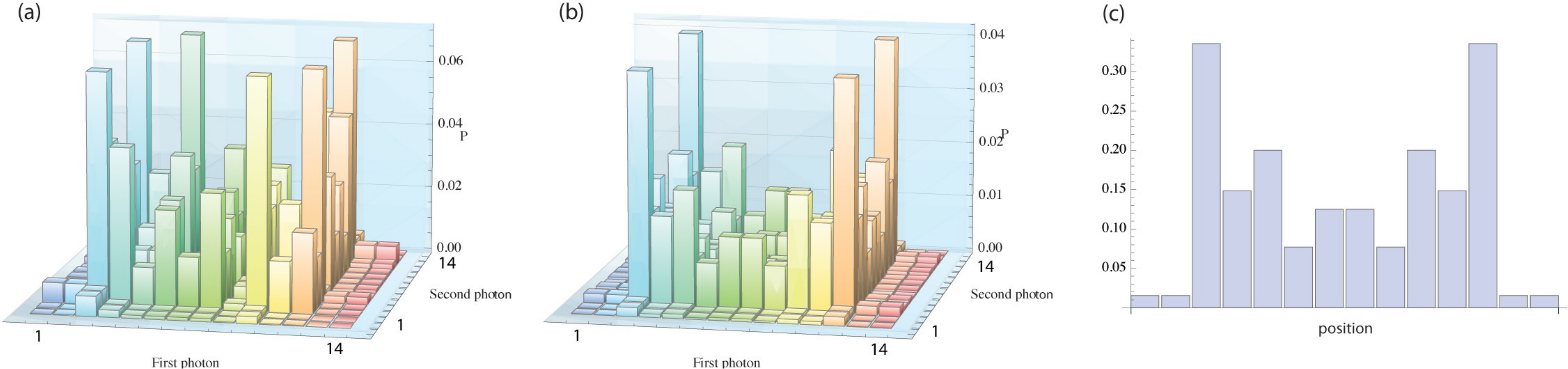}
\caption{(Color online). Two photon quantum walk with 7 levels. (a) Joint probability distribution with two indistinguishable walkers. (b) Joint probability distribution with two distinguishable walkers. (c) Single photon statistics for both distinguishable and indistinguishable walkers. Visually there is no discernable difference in (c) between the distinguishable probabilities ($P_i^{(1)}$) and indistinguishable probabilities ($P_i^{(2)}$), although there is a small numerical difference \mbox{$L_1 = \frac{1}{2} \sum_i |P_i^{(1)} - P_i^{(2)}| = 0.04$}. As with the 4 level system, in the case of distinguishable walkers the plot is uncorrelated, whereas in the case on indistinguishable walkers it is correlated.} \label{fig:theory_plots_7_levels}
\end{figure*}

We ran a simulation with both 4 levels and 7 levels in the tree, making 8 and 14 output modes respectively. In Figs. \ref{fig:theory_plots_4_levels} and \ref{fig:theory_plots_7_levels} we plot the probability amplitudes for the different terms in the output expression as a JPD. Let the numbers on the axes be denoted by $m$ and $n$. Then the element $\{m,n\}$ corresponds to the term with a photon at mode $m$ and a photon at mode $n$. The matrix is symmetric since there is no way of distinguishing $\{m,n\}$ from $\{n,m\}$. The diagonal terms correspond to the terms where both photons are in the same output mode, while the off-diagonal terms represent the terms where the two photons are in different output modes. This leads to a normalization condition for the JPD \footnote{In the case of indistinguishable photons, for normalization we require that the sum of the lower triangular elements of the matrix be 1, \mbox{$\sum_{i=1}^N\sum_{j=1}^{i} P_{ij}=1$},  where $N$ is the number of output modes. We only consider the lower triangular components because the symmetric components of the matrix are identical and correspond to the same terms in the output state. Thus they should only be counted once.}. This graph characterizes all the coincidence probabilities associated with the different terms in the output state. Note that all of the off-diagonal elements of the JPD can be experimentally obtained using coincidence measurements, which does not require number resolving detection. However, the diagonal terms, in which both photons appear at the same output (the meeting problem), inherently requires number resolving detection.

Next we consider the probability of detecting one or more photons at the $m$th mode, i.e we are now considering single click events rather than coincidence events. If we take the $m$th column from the previous matrix and sum its elements, we obtain the net probability of detecting a photon in mode $m$. This is shown in Fig. \ref{fig:theory_plots_7_levels}(c). This is analogous to the single walker case where we plot the probability distribution of the single photon statistics. Interestingly, the projection onto single click events in the case of indistinguishable and distinguishable walkers produce very similar statistics. The full matrix, however, contains much more information as it characterizes both single photon and coincidence statistics.

The two photon JPD and its projection onto one dimension can both be directly experimentally measured. The former can be measured by performing coincidence measurements, which tells us the probability associated with the respective terms, $\{m,n\}$. The projection can be measured using single (non-coincidence) measurements at the outputs.

Notice that in the JPD of indistinguishable particles exist clear oscillations between different terms, which are distinct from the distinguishable case. This structure is lost upon projection into one dimension. In the case of the 4 level system, it is evident that the two photon interference is characterized by the suppression of the two diagonal peaks. In the case of the distinguishable photons the JPD is uncorrelated, whereas in the indistinguishable case it is not.

An interesting question is, how do we define the `output' of the walk? Is it given by the single photon or the coincidence statistics? If we define the output as being the single photon statistics then the single and two walker situations are very similar -- they produce nearly identical spectra. However, if we define the output as being the set of coincidence statistics, then we have a much richer data set, which does not have a counterpart in the single walker scenario.



\subsection{The transition from distinguishable to indistinguishable walkers} \label{sec:int_indep}

The analysis has, up until this point, considered the situation of two indistinguishable walkers. However, in an arbitrary quantum walk, distinguishable walkers are also possible with an appropriate choice of coin and step operators. This arises naturally with distinguishable photons. We now turn our attention to comparing the situations where the walkers are indistinguishable and distinguishable, and examine the differences in the coincidence statistics in both scenarios. Let the input modes to the first beamsplitter be denoted 0 and 1. Then, when the incident photons are in the same spatio-temporal mode, the input state can be expressed $a_0^\dag a_1^\dag$. On the other hand, when the incident photons are distinguishable, we can express the input state as  $a_0^\dag b_1^\dag$, where $b$ denotes a mode orthogonal to $a$. In the former case the incident photons will interfere within the quantum walk, whereas in the later case they will not and instead evolve completely independently of one another and no interference will be observed. In the indistinguishable case we obtain JPDs given in Figs. \ref{fig:theory_plots_4_levels}(left) and \ref{fig:theory_plots_7_levels}(a), whereas in the distinguishable case the JPDs are illustrated in Figs. \ref{fig:theory_plots_4_levels}(right) and \ref{fig:theory_plots_7_levels}(b) \footnote{In the case of distinguishable walkers, each `slice' along one axis in the matrix corresponds to the single walker plot, weighted according to the respective element of the single walker plot on the other axis.}. Notice the differing interference patterns between the indistinguishable and distinguishable walker scenarios \cite{bib:Bromberg09}. However, upon projection onto single click events, both scenarios produce similar statistics. In the case of distinguishable walkers, the JPD is given by the product of the single click statistics along each axis, whereas in the case of indistinguishable walkers the JPD is correlated.

In the case of distinguishable walkers, $\{m,n\}$ and $\{n,m\}$ are now distinct. This leads to a different normalization condition than for the indistinguishable case \footnote{In the case of distinguishable photons, $\{m,n\}$ and $\{n,m\}$ are distinct (assuming the detectors can resolve the difference between $a^\dag$ and $b^\dag$). This results in the normalization condition, \mbox{$\sum_{i=1}^N\sum_{j=1}^N P_{ij}=1$}. i.e. we sum over all components in the matrix since each component is distinct from every other.}.

The two situations above, of indistinguishable and distinguishable walkers, represent the two extreme cases. In the in-between scenario whith an arbitrary degree of interference, we can decompose the two photon state into overlapping and non-overlapping components \cite{bib:RohdeMauererSilberhorn07} and represent the input state as  \mbox{$\alpha a_0^\dag a_1^\dag + \sqrt{1-\alpha^2} a_0^\dag  b_1^\dag$}, where $\alpha \in \Re$ quantifies to what extent the walkers are indistinguishable or distinguishable. In this case, the probabilities in the JPD are simply given by a linear combination of the two extreme cases \footnote{Following application of a unitary operator, which implements the beamsplitter network, the output state can be expressed in the form \mbox{$\sum_{ij} (\alpha \gamma_{ij} a_i^\dag a_j^\dag + \sqrt{1-\alpha^2} \xi_{ij} a_i^\dag b_j^\dag)$}. The net probability of detecting a photon in mode $i$ and a photon in mode $j$ is thus, \mbox{$P_{ij} = \alpha^2 |\gamma_{ij}|^2 + (1-\alpha^2) |\xi_{ij}|^2$}, assuming the detector is able to resolve the distinct modes $a$ and $b$. Thus the probability of detecting photons in modes $i$ and $j$ is simply a linear combination of the respective probabilities in the two extreme cases.}.

Note that there are two ways of interpreting the parameter $\alpha$. On one hand, with complete control over the system, it can be used to `tune' the extent to which the walkers are indistinguishable or distinguishable. On the other hand, it could be forced upon the experimenter as a result of mode-mismatch, time-jitter, or other forms of distinguishability between the incident walker photons.


\subsection{Photon correlation}

As noted above, the defining feature of walks with indistinguishable walkers is the correlation that exist between the photons in the system, which is not the case for distinguishable walkers. This arises as a result of interferometric suppression or enhancement of particular terms in the JPD. This was experimentally observed by Bromberg et. al in a continuous time quantum walk \cite{bib:Bromberg09}. More formally we can define a correlation measure as follows: (1) write the JPD in a matrix form, (2) diagonalize the JPD matrix with diagonal elements given by $p_i$, (4) renormalize the elements such that
\begin{equation}
p_i \to \frac{p_i}{\sum_j p_j}
\end{equation} 
gives a probability distribution, and (5) calculate the Shannon entropy of the resulting probability distribution
\begin{equation}
H(P) = -\sum_i p_i \, \mathrm{log} \, p_i .
\end{equation}

As an example we consider the 4 level walk presented earlier. For distinguishable walkers the JPD is uncorrelated, so it is diagonalizable with a single non-zero coefficient, hence $p_1 = 1$ and $H(P) = 0$. On the other hand, for indistinguishable walkers $H(P)$ is necessarily greater than zero. In this example we have $H(P) = 1.74$, a characteristic of photon correlations.


\subsection{Comparison with coherent state implementations}

We conclude our discussion of optical implementations of quantum walks by comparing to implementations where coherent states, rather than single photons, are employed. Such a system was recently demonstrated by Schreiber et. al \cite{bib:Schreiber10}.

We show that an arbitrary quantum walk system, in which all walkers are initialized into the same position state, can be simulated with coherent light (i.e. both will yield identical output statistics). This is consistent with previous observations in the continuous time regime \cite{bib:Bromberg09}. On the other hand, systems where walkers are initialized into different position states cannot, in general, be simulated by coherent light.

The evolution of a system containing $n$ steps, with $p$ walkers all initialized into position $x$, is given by
\begin{equation} \label{eq:Uadag}
 (U a_x^\dagger)^p\ket{0}= \Big(\sum_{k}{c_{k} a_{k}^\dagger\Big)^p\ket{0}},
\end{equation}
where $U=(SC)^n$. Let us now replace the $p$-photon Fock state with a coherent state $\ket{\alpha}=e^{-\frac{|\alpha|^2}{2}} e^{\alpha a_x^\dagger} \ket{0}$, also initialized at position $x$. The evolution is given by
\begin{eqnarray}
U e^{-\frac{|\alpha|^2}{2}} e^{\alpha a_x^\dagger} U^\dagger\ket{0} &=& e^{-\frac{|\alpha|^2}{2}} e^{U\alpha a_x^\dagger U^\dagger}\ket{0} \nonumber \\
&=& e^{-\frac{|\alpha|^2}{2}} e^{\sum_{k}\alpha c_{k} a_{k}^\dagger}\ket{0}
\end{eqnarray}
The second line follows from \ref{eq:Uadag}. Performing a Taylor series expansion we obtain
\begin{equation}
e^{-\frac{|\alpha|^2}{2}} e^{\sum_{k}\alpha c_{k} a_{k}^\dagger}\ket{0}= e^{-\frac{|\alpha|^2}{2}} \sum_d \frac{1}{d!}\Big(\sum_{k}\alpha c_{k} a_{k}^\dag\Big)^d\ket{0}
\end{equation}
If we now condition on $p$ walkers ($d = p$) (in the optical context using photon number resolving photo-detection) we obtain the same state as in the Fock state example, up to a constant factor $c$
\begin{equation}
\underbrace{e^{-\frac{|\alpha|^2}{2}} \frac{\alpha^p}{p!}}_c \Big(\sum_{k} c_{k} a_{k}^\dagger\Big)^p\ket{0}
\end{equation}
Since $c$ is constant it does not change the distribution of the single click or coincidence statistics and the distribution reflects that of the Fock state scenario.

On the other hand, if multiple photons are entering the system at different inputs, in general it is not possible to simulate the system using coherent light. A simple counter-example illustrates this. Consider the most trivial two mode quantum walk -- a single beamsplitter with two inputs and two outputs. Should we put a single photon into each beamsplitter port the system undergoes the transformation $\ket{11} \to (\ket{20}+\ket{02})/\sqrt{2}$. Thus upon performing number resolving detection, we will always measure both photons at one port or both photons at the other (each with 50\% probability), whereas one photon at each output will never be measured. With two coherent state inputs, there is no choice of phase relationships which can mimic these statistics. The defining difference here is that with single photons non-classical (HOM) interference is taking place.

More formally we can consider the general case where an incident coherent state $\ket{\alpha}$ propagates through an $N$-port linear network consisting only of beamsplitters and phase-shifters. Such a network transforms the incident state into a product state of different coherent states,
\begin{equation}
\ket{\alpha} \to \bigotimes_i \ket{\beta_i}
\end{equation}
Upon measuring single photons at output modes $m$ and $n$, and no photons at all other modes, we obtain
\begin{eqnarray}
\ket{\psi_\mathrm{out}} &=& e^{-|\beta_m|^2/2} \beta_m a^\dag_m . e^{-|\beta_n|^2/2} \beta_n a^\dag_n \prod_{i\neq m,n}e^{-|\beta_i|^2/2} \ket{0} \nonumber \\
&=& \underbrace{\beta_m a^\dag_m}_{f(m)} \underbrace{\beta_n a^\dag_n}_{f(n)} \underbrace{\prod_i e^{-|\beta_i|^2/2}}_{constant} \ket{0}
\end{eqnarray}
Note that the state is of a separable form. In our previous discussions we observed that walks consisting of two walkers at different inputs generate non-separable coincidence statistics, which is impossible in this case. Thus, in general, coherent states can only simulate multiple walkers in special cases: (1) with distinguishable photons, which generate separable statistics, and (2) the example presented above where multiple walkers begin in the same initial state. Importantly, even with quantum \emph{measurements}, such a system cannot mimic the statistics of a system with quantum \emph{states}.

\section{Conclusion} \label{sec:conclusion}

We have presented a framework for describing multi-walker, discrete time quantum walks on arbitrary graphs, considering the cases of both distinguishable and indistinguishable walkers. We presented several example applications for this framework on different graph structures and discussed the specific example of photonic quantum walks. We discussed measurement and characterization of quantum walks and the issue of measurement correlations, which differentiate distinguishable versus indistinguishable walkers. We have shown that single walker walks have linear scaling in terms of the number of parameters describing the system and that the complexity of the system scales exponentially with the number of walkers. Quantum walks have been shown to be universal for quantum computation, and the framework we have presented provides the generic tools necessary to describe this.


\begin{acknowledgments}
We acknowledge Andreas Eckstein, Malte Avenhaus, Aur\'el G\'abris and V\'aclav Poto\v cek for very helpful discussions. In particular we thank Andreas Christ and Katiuscia Cassemiro for immeasurably helpful discussions and critique. PR acknowledges support from the Australian Research Council.  M.S and I.J acknowledge the grant support from the Czech Ministry of Education MSM6840770039 and MSMT LC06002. We acknowledge the German Israel Foundation (Project 970/2007) for financial support. While preparing this manuscript Peruzzo et. al \cite{bib:Peruzzo10} experimentally demonstrated an optical implementation of a two photon quantum walk which overlaps with the techniques presented here.
\end{acknowledgments}


\bibliography{bibliography.bib}

\end{document}